\documentclass[12pt]{article}

\usepackage{newtxtext,newtxmath}

\usepackage[version=4]{mhchem}
\usepackage{graphicx}

\usepackage{lineno}

\usepackage[letterpaper,margin=1in]{geometry}

\renewenvironment{abstract}
	{\quotation}
	{\endquotation}

\date{}

\makeatletter
\renewcommand{\fnum@figure}{\textbf{Figure \thefigure}}
\renewcommand{\fnum@table}{\textbf{Table \thetable}}
\makeatother

\usepackage{scicite}

\usepackage{url}

\newcommand{\kct}{\ce{K_{1-x}CrTe2}}
\newcommand{\ksf}{\ce{K_{0.64}CrTe2}}

\def\scititle{In-Plane Ferromagnetism and Critical Dynamics in Alkali-Deficient \ce{K_{1-x}CrTe2} (with $x \approx$ 0.3) Single Crystals
}
\title{\bfseries \boldmath \scititle}

\author
{Catherine Witteveen,$^{1,2\ast}$ 
Felix Eder,$^{1}$  
Sara A. L\'{o}pez-Paz,$^{3}$\and
Vladimir Pomjakushin,$^{4}$ 
Jonas A. Krieger,$^{4}$ 
Zurab Guguchia,$^{4}$ 
Harald O. Jeschke,$^{5}$\and 
Martin M\aa nsson,$^{6}$
Fabian O. von Rohr$^{1}$\and 
\small$^{1}$Department of Quantum Matter Physics, University of Geneva, 1211 Geneva, Switzerland.\and
\small$^{2}$Leibniz-Institut für Festkörper u. Werkstoffforschung, 01069 Dresden, Germany.\and
\small$^{3}$Department of Chemistry, University of Copenhagen, 2100 Copenhagen, Denmark.\and
\small$^{4}$PSI Center for Neutron and Muon Sciences, 5232 Villigen, Switzerland.\and 
\small$^{5}$Research Institute for Interdisciplinary Science, Okayama University, 700-8530 Okayama, Japan.\and 
\small$^{6}$Department of Applied Physics, KTH Royal Institute of Technology, 06 91 Stockholm, Sweden.\and
\small$^\ast$Corresponding author. Email: c.witteveen@ifw-dresden.de
}

\date{}

\begin{document} 

% Insert the title and author list
\maketitle

% Abstract, in bold
% There are strict length limits, and not all formats have abstracts.
% Consult the journal instructions to authors for details.
% Do not cite any references in the abstract.

\begin{abstract} \bfseries \boldmath
% Start with one or two sentences of background
Layered chromium tellurides are model systems for studying low-dimensional magnetism in van der Waals materials. We report the synthesis and characterization of \ce{K_{1-x}CrTe2} single crystals ($x \approx 0.3$), which crystallize in the \textit{Cm} space group with trigonal prismatic \ce{K^{+}} coordination, unlike the octahedral environments of more stoichiometric \ce{ACrX2} compounds. Magnetization measurements show a sharp ferromagnetic transition at $T_{\rm C}=117$ K and in-plane magnetic anisotropy, supported by density functional theory. Neutron diffraction reveals ferromagnetic alignment of Cr spins within and between layers. This contrasts with the out-of-plane A-type antiferromagnetism in \ce{LiCrTe2} and \ce{NaCrTe2}, but resembles \ce{CrTe2}. These differences likely arise from changes in interlayer spacing, Cr oxidation state, or stacking. AC susceptibility and $\mu$SR indicate short-range order above $T_{\rm C}$ and dynamic behavior below. Overall, \ce{K_{1-x}CrTe2} provides a tunable platform for studying spin orientation and dimensionality in two-dimensional magnets.

\end{abstract}

%\linenumbers
% In setting up this template for *Science* papers, we've used both
% the \section* command and the \paragraph* command for topical
% divisions.  Which you use will of course depend on the type of paper
% you're writing.  Review Articles tend to have displayed headings, for
% which \section* is more appropriate; Research Articles, when they have
% formal topical divisions at all, tend to signal them with bold text
% that runs into the paragraph, for which \paragraph* is the right
% choice.  Either way, use the asterisk (*) modifier, as shown, to
% suppress numbering.

\section*{Introduction}

Alkali-metal intercalated transition-metal dichalcogenides have long been studied for energy storage applications, but they also exhibit a rich variety of correlated electronic and magnetic phenomena. A prototypical example is \ce{Na_xCoO2}, originally developed in the 1980s as a solid-state cathode for sodium--ion batteries.\cite{shacklette1988rechargeable} Beyond its electrochemical relevance, \ce{Na_xCoO2} has been found to display exceptional thermoelectric performance,\cite{wang2003spin,terasaki1997NaCoO2} complex vacancy-driven charge and magnetic orderings,\cite{foo2004charge,roger2007patterning,schulze2008direct,medarde20131d} and even superconductivity in its hydrated form at $x \approx$ 0.3,\cite{takada2003,schaak2003superconductivity} providing a rich system for exploring how ionic ordering and electronic correlations govern emergent phases in low-dimensional delafossite-type materials.

The structurally related, partially intercalated chromium dichalcogenides \ce{\textit{A}_xCr\textit{X}2} (\textit{A} = monovalent cation, \textit{X} = dichalcogenide) form a family of layered materials where magnetic Cr\textsuperscript{III}/Cr\textsuperscript{IV} ions occupy a triangular lattice, offering a fertile platform for exploring geometric frustration and competing exchange interactions in high-spin systems. In this regard, they share several conceptual similarities with the well-studied \ce{Na_xCoO2} system, including tunable cation content, charge ordering phenomena, and the emergence of complex magnetic ground states from a triangular net structure.\cite{toth2016electromagnon,sugiyama2009li,nocerino2023competition} \textit{A}$_x$\ce{Cr\textit{X}2} compounds provide a versatile platform where magnetic interactions can be systematically tuned through variations in cation size and cation occupancy.\cite{kobayashi2019linear,song2021kinetics,song2019soft,kobayashi2016competition,xu2020intrinsic,huang2022anisotropic} Notably, recent studies have uncovered long-range magnetic order in undulated, incommensurately modulated K$_{1-x}$CrSe$_2$ ($x \approx$ 0.13), with an enhanced N\'{e}el temperature compared to its fully intercalated parent compound, as well as demonstrated stoichiometry and phase control by deliberate alteration of parameters of the synthesis protocol.\cite{eder2025,eder2025stoichiometry} 

The magnetic ground states across the \textit{A}$_x$\ce{Cr\textit{X}2} series result from a balance between competing exchange interactions, which can be well-described by the Goodenough-Kanamori-Anderson rules (GKA). In these layered systems, the high-spin Cr$^{3+}$ ions (3$d^3$) occupy a triangular lattice and interact through both the (i) direct Cr--Cr antiferromagnetic exchange, which dominates at short distances, and the (ii) Cr--X--Cr superexchange, which becomes ferromagnetic when the bond angle is getting close to 90$^\circ$.\cite{1995Lafond} The relative strengths of these interactions are highly sensitive to structural parameters such as bond angles and TM bond lengths, interlayer spacing, and local distortions. As a result, the \ce{\textit{A}Cr\textit{X}2} series hosts a broad range of magnetic ground states from a 120$^\circ$ spin structure (\ce{LiCrS2}), to a helical incommensurate structure (\ce{LiCrSe2}, \ce{NaCrS2}) up to an out-of-plane A-type AFM order (\ce{LiCrTe2}, \ce{NaCrTe2})\cite{engelsman1973,witteveen2023synthesis,huang2022anisotropic}, while deintercalated \ce{CrTe2} is a ferromagnet with in-plane spin orientation.\cite{Freitas2015,roseler2025efficient} Combined with the ability to grow large, high-quality single crystals via mixed \textit{A}/\textit{X} flux synthesis,\cite{witteveen2023synthesis} this family provides a platform to investigate how crystal structure and dimensionality govern magnetic interactions in triangular-lattice low-dimensional materials.

Here, we present the synthesis and a detailed characterization of the physical properties of \ce{K_{1-x}CrTe2}. Although \ce{KCrTe2} has previously been used as a precursor for the oxidative deintercalation route to ferromagnetic \ce{CrTe2} \cite{Freitas2015, roseler2025efficient}, its intrinsic structural and magnetic properties have remained vastly unexplored probably due to the encounter of numerous challenges in the determination of the crystal structure.\cite{prasad2026KCrTe2} Using a mixed K/Te flux method, we obtained high-quality single crystals of \ce{K_{1-x}CrTe2}, enabling a detailed structural and magnetic analysis by combining temperature-dependent single crystal X-ray diffraction (SXRD) measurements, density functional theory (DFT) calculations, neutron scattering, static and dynamic magnetization measurements as well as muon spin spectroscopy ({$\mu$}SR). Our measurements reveal that \ce{K_{1-x}CrTe2} undergoes magnetic ordering below $T_{\rm C} \approx$ 117 K, with an in-plane spin orientation similar to that of deintercalated \ce{CrTe2}. This magnetic anisotropy contrasts with the out-of-plane easy axes observed in Li- and Na-based \ce{\textit{A}CrTe2} analogs, suggesting that this compound lies near a crossover regime in the series’ magnetic phase diagram. A comparison of the structural parameters taken by SXRD and the magnetic properties of the members of this series is given in Table \ref{table1} (and Table \ref{SM_tab2} in the SM). The values for \ce{K_{1-x}CrTe2} are highlighted in blue, which shows two different in-plane Cr--Cr distances ($d_{\rm intra}$) and  two different Cr--Te--Cr angles ($\alpha$). Our findings demonstrate that \ce{K_{1-x}CrTe2} serves as a bridge between A-type antiferromagnetic and ferromagnetic ordering, offering new insight into how intercalated alkali content, structural modulation, and exchange geometry collectively govern magnetism in triangular-lattice chromium tellurides.

\begin{table}[h] 
\small
  \caption{\textbf{Comparison of the \textit{A}CrTe$_2$ series.} Structural parameters are taken from SXRD measurements at the respective temperatures. $d_{\rm intra}$ is the intralayer and $d_{inter}$ the interlayer Cr--Cr distance while $\alpha$ is the Cr--Te--Cr angle. Data from \ce{CrTe2} originates from reference \cite{roseler2025efficient}. In blue is highlighted the compound object of this communication.}
  \label{tab_Ch6}
  \begin{tabular*}{\textwidth}{@{\extracolsep{\fill}} c c c c c c c}
    \hline
      &T (K)& $d_{\rm intra}$ (\text{$\mathring{\mathrm{A}}$}) & $\alpha$ &$d_{inter}$ (\text{$\mathring{\mathrm{A}}$})&Magnetic structure & $T_{\rm T}$ (K) \\
    \hline
     \ce{CrTe2} & 120  & 3.7823 & 89.77° & 6.0203 & in-plane FM & 318  \\
    \ce{LiCrTe2} & 90  & 3.9673(1) & 92.854(8)°  & 6.5876(1) & out-of-plane AFM & 148   \\
     \ce{NaCrTe2} & 100  & 3.9979(1) & 93.404(7)°  & 7.3735(2)  & out-of-plane AFM & 110  \\
   \color{blue} \ce{K_{0.64(4)}CrTe2} & \color{blue} 100 &\color{blue}  3.9466(3) &\color{blue}  92.3(2)°  & \color{blue} 8.3714(7) & \color{blue} in-plane FM & \color{blue} 117 \\
    & & \color{blue} 3.9112(2) & \color{blue} 91.6(2)°  & &  & \\
    \hline
  \end{tabular*}
  \label{table1}
\end{table}

\section*{Results}

\subsection*{Crystal growth and structural characterization of \ce{K_{1-x}CrTe2}}

Single crystals of \ce{K_{1-x}CrTe2} were grown using a K/Te flux, (see the Materials and Methods section of the Supplementary Materials (SM) for the detailed procedure).\cite{methods} This method was adapted from the previously reported synthesis of large LiCrTe$_2$ single crystals.\cite{witteveen2023synthesis} After centrifugation at high temperature, crystals with dimensions up to 6 mm x 6 x mm x 0.5 mm and a silvery metallic luster were extracted. The crystals were found to be extremely sensitive to air. Energy dispersive X-ray spectroscopy (EDX) done on crystals of different batches resulted in variations of the K content between 0.66(11) and 0.78(5) per Cr atom (details of the EDX results can be found in Table \ref{SM_tab1} of the SM).
By means of single-crystal X-ray diffraction, the crystal structure of \ce{K_{1-x}CrTe2} could be solved in the monoclinic \textit{Cm} space group with lattice parameters (100 K) of \textit{a} = 6.7539(4) $\mathring{\mathrm{A}}$, \textit{b} = 3.9466(3) {$\mathring{\mathrm{A}}$}, \textit{c} = 8.6549(7) {$\mathring{\mathrm{A}}$} and $\beta$ = 104.705(8)° with a cell-volume of 223.14(3) $\mathring{\mathrm{A}}$$^3$ and a refined potassium content of 0.64(4) K per Cr, in line with the variations observed in EDX. The crystallographic data is listed in Table \ref{SM_tab2} in the SM.

\begin{figure*}
\begin{center}
\includegraphics[width=\linewidth]{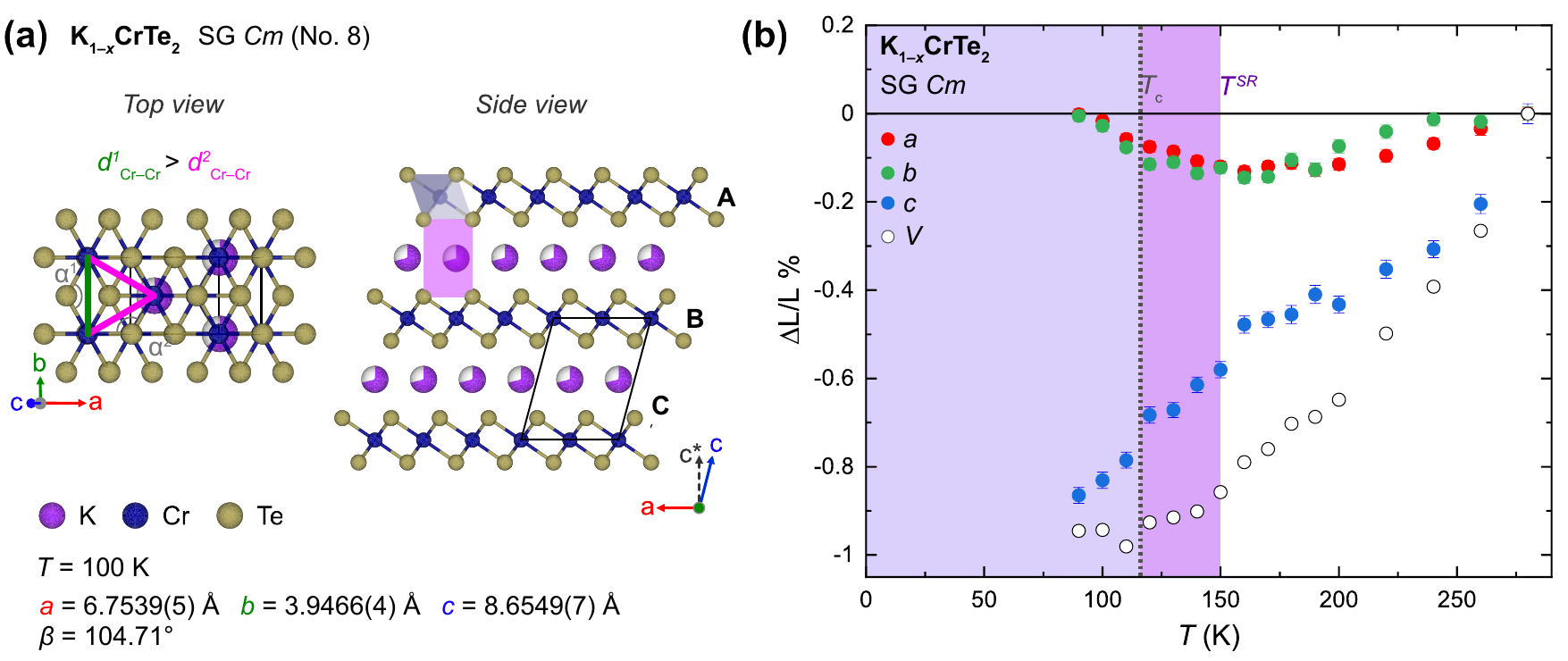}
\caption{\textbf{Crystal structure of \ce{K_{1-x}CrTe2}}. (a) Schematic representation of the crystal structure in a projection on the (001) plane (left) and along the [010] axis (right). K cations are drawn purple, Cr atoms blue and Te atoms yellow. The triangular lattice is not equilateral, with  \textit{b} being shorter. (b) Temperature dependence of lattice parameters \textit{a}, \textit{b}, \textit{c} and the cell volume \textit{V} determined from SXRD measurements in a temperature range between 90 K and 280 K normalized to 280 K. }
\label{KCrTe2_Fig1}
\end{center}
\end{figure*}

In Fig. \ref{KCrTe2_Fig1} (a) we show the top (along \textbf{c$^*$}) and side view (along \textbf{b}) of the crystal structure of \ce{K_{1-x}CrTe2}, which consists of \ce{CrTe2} layers intercalated by K cations. The Cr atoms form a triangular lattice within the \ce{CrTe2} layers and are located at the center of octahedra, which are slightly trigonally distorted and composed of tellurium ions. The in-plane \textit{d}$_{Cr-Cr}$ distances are not equilateral, as is shown by the triangle in Fig. \ref{KCrTe2_Fig1} (a). While ${d_1}$$_{Cr-Cr}$ (green) is equal to the \textit{b} lattice parameter, ${d_2}$$_{Cr-Cr}$ (pink) is slightly smaller (3.9466(4) {$\mathring{\mathrm{A}}$} vs 3.9112(2) {$\mathring{\mathrm{A}}$} at 100 K). This is also impacting the Cr--Te--Cr angles. Along the $<$010$>$ directions the angles are slightly higher ($\alpha_1$ = 92.3(2) and 92.4(2)°) than along the $<$110$>$ directions ($\alpha_2$ = 91.2(2) and 91.6(2)°). The \textit{ABCABC} layer stacking is shared with rhombohedral \ce{KCrS2} and pseudo-rhombohedral \ce{KCrSe2}.
While in stoichiometric \ce{\textit{A}Cr\textit{X}2} compounds with \textit{A} = Li, Na, K the alkali cations are coordinated octahedrally, in \ce{K_{1-x}CrTe2}, the K cations exhibit a trigonal prismatic environment.\cite{Nocerino2022, Kobayashi2016, nocerino2023competition,engelsman1973,Song2021,eder2025,Ushakov2013} A similar coordination was reported for the likewise under-stoichiometric compounds Na$_{\sim0.6}$CrSe$_2$, K$_{0.6-0.8}$CrSe$_2$ and $\beta$--Na$_{0.67}$CoO$_2$, while in the incommensurately modulated Na$_{\sim0.78}$CrO$_2$, both coordinations as well as intermediate forms are observed.\cite{2002Ono, Miyazaki2021} This different coordination environment has been denoted as "type I" for the trigonal prismatic, and "type II" for the octahedral coordination in the past.\cite{VanBruggen1978, Wiegers1980, Nikiforow1999}

%\subsubsection*{Negative Thermal Expansion}

Our temperature-dependent SXRD measurements of \ce{K_{1-x}CrTe2} reveal a striking anisotropy in thermal expansion between 90 and 280 K as shown in Fig. \ref{KCrTe2_Fig1} (b). In layered magnetic materials, thermal expansion behavior often provides insight into underlying spin–lattice interactions and electronic instabilities. Negative thermal expansion (NTE), in particular, has been linked to spin fluctuations in several van der Waals magnets. While \textit{c} expands monotonically, the in-plane lattice parameters (\textit{a} and \textit{b}) contract upon warming up to $\sim 150$ K, which we will denote here tentatively as the temperature where magnetic short-range order ends (\textit{T}$^{SR}$), with a thermal expansion coefficient of $\alpha_{\textit{a},\textit{b}}$ (80-150 K) $= -1.97 \times 10^{-5}$ K$^{-1}$. Such behavior resembles other vdW magnets like CrSBr or \ce{LiCrTe2}, where continuous in-plane NTE is closely tied to magnetic order.\cite{Lopez-Paz2022, Nocerino2022} Our neutron powder diffraction (NPD) refinements over the same temperature range show no evidence of a structural phase transition, confirming that the lattice symmetry remains intact throughout. 

In \ce{K_{1-x}CrTe2}, the in-plane NTE terminates around 150 K -- above the magnetic ordering temperature $T_{\rm C} = 117$ K -- and coincides with the onset of dynamic magnetic fluctuations observed in AC susceptibility (see the discussion below). This decoupling of NTE from long-range magnetic order, yet alignment with emergent spin dynamics, suggests that local spin--lattice interactions may play a key role in driving structural anomalies in the absence of a phase transition.

\subsection*{Density functional theory calculations}

\begin{figure*}
\begin{center}
\includegraphics[width=\linewidth]{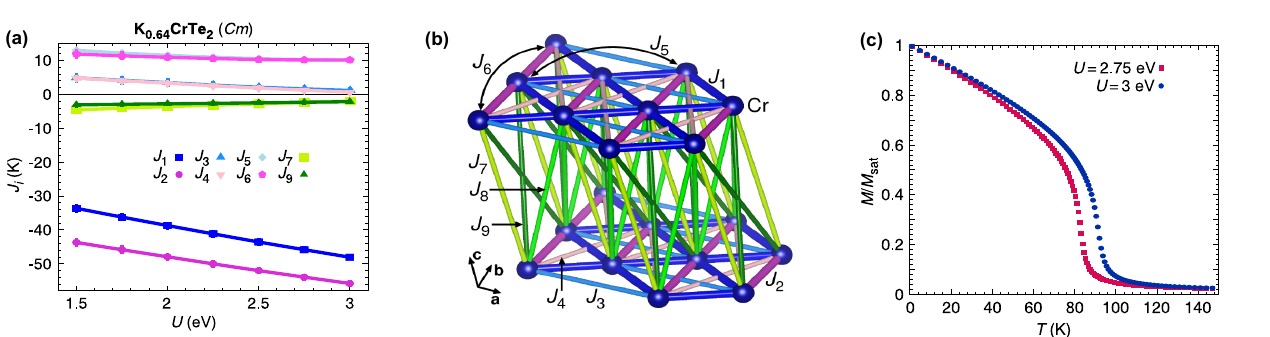}
\caption{{\bf Calculation results for the Heisenberg Hamiltonian of \ce{K_{0.64}CrTe2}.} (a) Eight exchange interactions as function of the on-site interaction strength $U$ from the DFT energy mapping result. (b) Geometry of relevant in-plane and interlayer exchange paths. (c) Magnetization curves for  \ce{K_{0.64}CrTe2} calculated with classical Monte Carlo, plotted relative to the saturation magnetization $M_{\rm sat}$. The two chosen sets of Heisenberg Hamiltonian parameters yield ferromagnetic transition temperatures of $T_{\rm C}=83$\,K and $T_{\rm C}=91$\,K, respectively.}
\label{KCrTe2_Fig2}
\end{center}
\end{figure*}

We investigate the magnetic interactions in {\kct} using density functional theory calculations with the full potential local orbital basis~\cite{Koepernik1999} and a generalized gradient approximation exchange correlation functional~\cite{Perdew1996}. We deal with the strong electronic correlations on the Cr $3d$ orbitals using a DFT+U correction~\cite{Liechtenstein1995}. We deal with the potassium deficiency in the stoichiometry \ce{K_{0.64}CrTe2} by applying the virtual crystal approximation (VCA)~\cite{Nordheim1931} which allows us to tune the nuclear charge of the alkali site between potassium and argon. This approach simulates a perfectly random distribution of reduced positive charge in the alkali layer. Applying VCA in magnetic materials with partial occupations has yielded very good results in \ce{Lu2Mo2O5N2}~\cite{Iqbal2017} or Y-kapellasite~\cite{Hering2022}, for example. We determine the parameters of the Heisenberg Hamiltonian 
\begin{equation} \label{eqDFT}
H = \sum_{i<j} J_{ij}\, \mathbf S_i \cdot \mathbf S_j \, ,
\end{equation}
where $\mathbf S_i$ are $S=3/2$ spins for Cr$^{3+}$. For this purpose, we apply the DFT energy mapping technique that has yielded excellent results for various Cr based magnets~\cite{Ghosh2019,Guo2024,Jaubert2025,Nilsen2025} and in particular also for layered Cr tellurides~\cite{witteveen2023synthesis,roseler2025efficient}. We focus on the {\ksf} structure with \textit{Cm} space group described above. We create a 16-fold supercell of the average structure with 16 inequivalent Cr spins that allows us to extract the values of the first nine exchange interactions shown in Fig.~\ref{KCrTe2_Fig2}\,(b) except for $J_8$. This allows us to assess both the in-plane ($J_1$ to $J_6$) and the interlayer ($J_7$ to $J_9$) interactions. The DFT+U calculations have two parameters, the Hund's rule coupling which we fix to a literature value of $J_{\rm H}=0.72$\,eV~\cite{Mizokawa1996} and the on-site interaction $U$ which we vary. The calculated exchange couplings are shown in Fig. ~\ref{KCrTe2_Fig2}\,(a) for seven values of $U$. The two nearest-neighbor couplings constituting the anisotropic triangular lattice of {\ksf} are strongly ferromagnetic. Second and third neighbors in the plane are much smaller and antiferromagnetic, somewhat destabilizing the ferromagnetic layers. The two interlayer couplings $J_7$ and $J_9$ are small and ferromagnetic. Resolving the most likely also small interlayer coupling $J_8$ would require calculations for larger supercells and therefore much larger computational effort. We also performed fully relativistic calculations of the ferromagnetic state with different quantization axes in order to extract the single ion anisotropy. We find \textbf{c} to be the hard axis. In space group \textit{Cm}, \textbf{a} and \textbf{b} directions are inequivalent, and there is a weak preference for the moments to point along \textbf{b}. 

We now proceed to show that the calculated Hamiltonian parameters lead to the correct ground state by performing classical Monte Carlo simulations. For two values of the on-site interaction strength $U$, we find the magnetization curves shown in Fig. ~\ref{KCrTe2_Fig2}\,(c). The ground state is ferromagnetic in agreement with experiment. The calculated ordering temperatures are below the experimental values. However, in layered magnets like {\kct}, it is well known that due to Mermin-Wagner physics, the ordering temperature has a logarithmic dependence on both interlayer coupling and single ion anisotropy~\cite{Yasuda2005}. As these are both small quantities, it is a hard problem to predict the exact ordering temperature of such a layered ferromagnet, but our results qualitatively reproduce the observed experimental behavior.

\subsection*{Neutron Powder Diffraction}

\begin{figure*}
\begin{center}
\includegraphics[width=\linewidth]{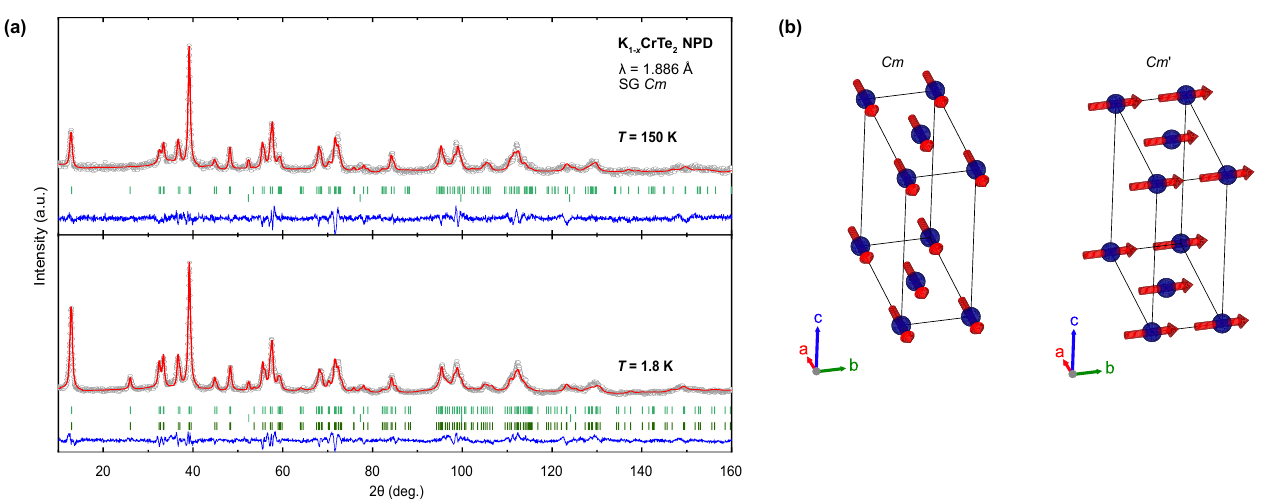}
\caption{\textbf{Long-range magnetic order of \ce{K_{1-x}CrTe2}.} (a) Rietveld refinement of the NPD data taken at 150 K and 1.8 K. In gray are the observed and in red the calculated intensities. The Bragg reflections are shown for the structural and magnetic phase, as well as for a secondary phase which belongs to vanadium from the sample container. The difference plot is drawn in blue. (b) The two possible solutions for the magnetic structure, showing the spins to lie in-plane.}
\label{KCrTe2_Fig3}
\end{center}
\end{figure*}

Temperature-dependent NPD revealed a ferromagnetic ground state with an in-plane easy axis for \ce{K_{1-x}CrTe2}. Fig. ~\ref{KCrTe2_Fig3}(a) shows Rietveld refinements of the NPD patterns collected at 150 K and 1.8 K. Both patterns are well fitted by the non-centrosymmetric monoclinic space group \textit{Cm}, with no evidence of a structural phase transition across the Curie temperature. This supports the structural solution obtained from single-crystal X-ray diffraction and confirms that the observed negative thermal expansion arises without symmetry breaking.

The lattice response is strongly anisotropic. Upon cooling from 150 K to 1.8 K, the out-of-plane \textit{c} lattice parameter contracts, while in-plane \textit{a} and \textit{b} lattice parameters expand. This anisotropy leads to a shortening of Cr--Te, K--Te, and Te--Te distances along the stacking direction, and an increase in in-plane Cr--Cr separations. These findings are in agreement with the thermal expansion behavior observed in SXRD and highlight a pronounced coupling between lattice parameters and magnetism.

In the base temperature diffraction pattern at 1.8 K we observe additional scattering intensity at nuclear positions, consistent with a ferromagnetic structure characterized by a propagation vector $\mathbf{q} = (0, 0, 0)$. Magnetic symmetry analysis yields two equally probable irreducible representations — \textit{Cm} and \textit{Cm}' — each supporting in-plane alignment of the Cr\textsuperscript{III}/Cr\textsuperscript{IV} moments along either \textit{a} or \textit{b} with no distinct inclination (Fig. ~\ref{KCrTe2_Fig3}(b)). The refined ordered moment reaches 2.73~$\mu_B$ per Cr at 1.8 K, in excellent agreement with the saturation magnetization obtained from DC measurements, and consistent with the expected value for $x \approx 0.3$.

%Finally, the temperature evolution of the refined magnetic moment within the b-axis (\textit{Cm}'), as suggested from DFT to be the slightly preferred alignment, follows a power-law behavior, yielding a Curie temperature of $T_{\rm C} = 117.8(5)$K and a critical exponent $\beta = 0.205(20)$ (Figure\ref{KCrTe2_Fig3}(c)). This $\beta$ value is close to the theoretical prediction for the two-dimensional XY universality class ($\beta \approx 0.231$), supporting the presence of easy-plane anisotropy and low-dimensional magnetic behavior in K$_x$CrTe$_2$.

\subsection*{Static and Dynamic magnetization measurements}

\begin{figure*}
\begin{center}
\includegraphics[width=\linewidth]{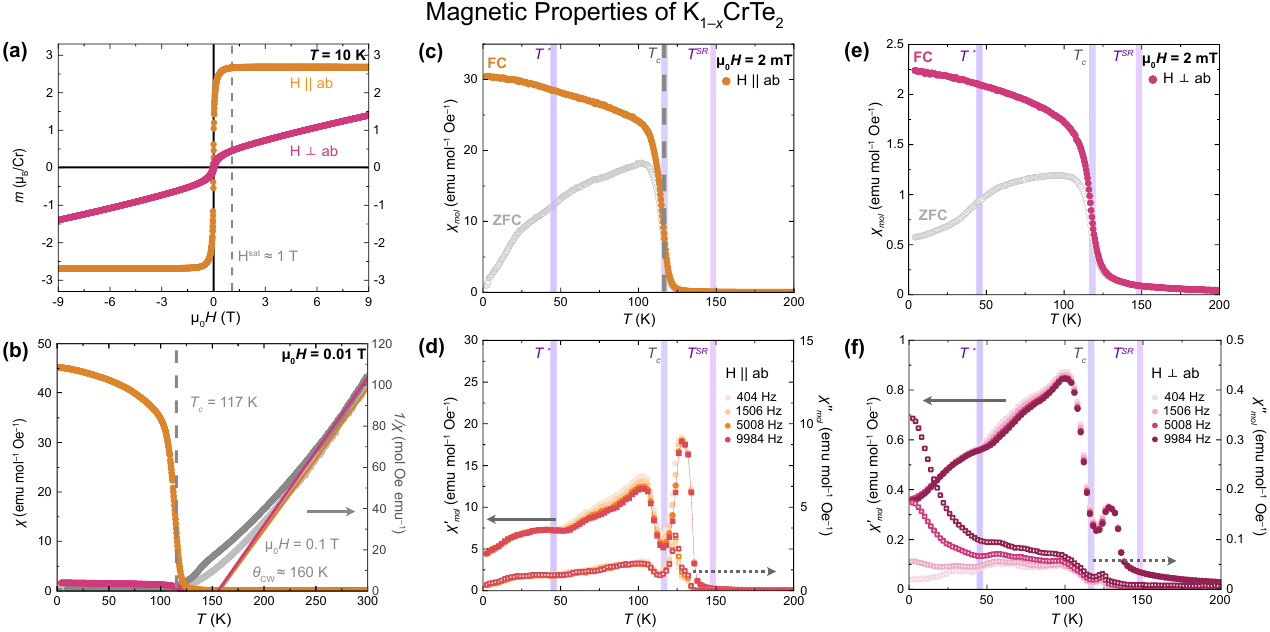}
\caption{\textbf{DC and AC magnetization of the \ce{K_{1-x}CrTe2} single crystals in both orientations: \textbf{H} $\parallel$ ab (orange) and \textbf{H} $\perp$ ab (pink).} (a) Field-dependent magnetic moment for both orientations taken at 1.8 K with a saturation of the moment at $\approx$ 1 T for \textbf{H} $\parallel$ \textbf{ab} (dashed line). (b) Left axis: FC molar susceptibility for both orientations at $\mu_0 H =$ 0.01 T. Right axis: the inverse susceptibility of both orientations taken at $\mu_0 H =$ 0.1 T showing $\Theta_{CW} \approx 160$ K. (c, e) \textit{T}-dependent molar susceptibility for both orientations at $\mu_0 H =$ 2 mT. (d,f) \textit{T}-dependent real (left axis) and imaginary component (right axis) of the AC susceptibility for both orientations. The shaded regions indicate the critical temperatures (\textit{T}$^{SR}$ = 150 K, \textit{T}$_{\rm C}$ = 117 K, \textit{T}$^*$ $\approx$ 50 K) where the phase transitions occur.}
\label{KCrTe2_Fig4}
\end{center}
\end{figure*}

Direction-dependent magnetization measurements reveal a ferromagnetic ground state in \ce{K_{1-x}CrTe2} with strong in-plane magnetic anisotropy. Isothermal $M(H)$ curves as shown in Fig. ~\ref{KCrTe2_Fig4}(a) show clear saturation for \textbf{H} $\parallel$ \textbf{ab} at $\sim$1 T, reaching a moment of 2.69 $\mu_B$/Cr, consistent with a mixed Cr$^{3+}$/Cr$^{4+}$ state for $x = 0.3$. In contrast, no saturation is observed up to 9 T for \textbf{H} $\perp$ \textbf{ab}, confirming the easy axis to lie in-plane. The absence of hysteresis points to soft ferromagnetism, reminiscent of ambient pressure \ce{CrSiTe3}~\cite{Zhang2021}.

Temperature-dependent susceptibility $\chi_{\text{mol}}(T)$ as shown in Fig. ~\ref{KCrTe2_Fig4}(b) shows ferromagnetic ordering, with a positive Curie–Weiss temperature of $\Theta_{\rm CW} \approx 160$ K indicating dominant ferromagnetic interactions as predicted by the GKA rules. The transition to a long-range magnetically ordered state at $T_{\rm C} = 117$ was determined by the first derivative of the temperature-dependent molar magnetic susceptibility taken at low fields, see Figs. ~\ref{KCrTe2_Fig4}(c, d) and Fig. \ref{Si_mag_derivative}. Deviations from Curie–Weiss behavior below $\sim$250 K, including a slope change in $\chi^{-1}(T)$ for \textbf{H} $\perp$ \textbf{ab} between 125--150 K, suggest significant short-range correlations already above the ordering temperature. This, combined with the anisotropic lattice response, non-equilateral triangular geometry and the calculated AFM couplings J5 and J6, could lead to possible magnetic frustration or slow spin dynamics. These signatures suggest a complex interplay between lattice geometry and magnetic interactions, motivating a closer look at the system’s dynamic response near the transition.

To probe the nature of spin dynamics in K$_{1-\textit{x}}$CrTe$_2$, we performed AC magnetic susceptibility measurements $\chi(\omega) \equiv dM/dH = \chi' + i\chi''$, for both in-plane and out-of-plane orientations, as shown in Figs. ~\ref{KCrTe2_Fig4}(d, f). The left axis depicts the real part and the right axis the imaginary part of the AC susceptibility.
%Several features point toward underlying magnetic frustration: (i) the non-equilateral triangular Cr\textsuperscript{III}/Cr\textsuperscript{IV} lattice, with two distinct Cr--Cr distances (${d_1}_{\text{Cr–Cr}} = 3.9466(4) ~\mathring{\mathrm{A}}$, ${d_2}_{\text{Cr–Cr}} = 3.9112(2)~\mathring{\mathrm{A}}$) and corresponding Cr--Te--Cr bond angles ($\alpha^1 = 92.3(3)^\circ$, $\alpha^2 = 91.2(2)^\circ$ at 100 K), leading to inequivalent superexchange paths; (ii) the anomalous negative thermal expansion persisting well above $T_{\rm C}$ ($T^{SR}$); (iii) a frustration index $f = \Theta_{\text{CW}}/T_{\rm C} \approx 1.34$; and (iv) clear deviations from Curie–Weiss behavior below $\sim$250 K. These signatures suggest a complex interplay between lattice geometry and magnetic interactions, motivating a closer look at the system’s dynamic response near the transition.

%In Fig. \ref{KCrTe2_Fig4} we show (c) zero field cooled (ZFC) and field cooled (FC) susceptibility taken with a DC field of 2 mT and in Fig. \ref{KCrTe2_Fig4} (d) AC susceptibility (real left axis, imaginary right axis) with an applied excitation AC field of 1 mT for the $K_{1-\textit{x}}$CrTe$_2$ crystals for the external field applied along the plane. In Fig. \ref{KCrTe2_Fig4} (e) and (f) the same for the field applied perpendicular to the plane. 
Three regions are distinguished: (i) The first starting at $\approx$ $T^{SR}$ = 150 K -- in line with the observation of the negative thermal expansion of the unit cell parameter \textit{a} and \textit{b} of the \textit{T}-dependent SXRD measurements (see Fig. ~\ref{KCrTe2_Fig1} (b)) -- until we reach our $T_{\rm C}$ at 117 K. There, a peak only observable in the AC susceptibility (Figs. ~\ref{KCrTe2_Fig4} (d, f)) emerges that shows no frequency-dependent temperature shift. The real (128 K) and imaginary (124 K) parts do not have the same maximum temperature in both orientations. (ii) The second appears below the $T_{\rm C}$ of 117 K up to $T^{*}$ with a stronger frequency dependence. (iii) The third lies below $T^*$ $\approx$ 50 K, where the real part of the susceptibility becomes frequency-independent again.

From Fig. \ref{KCrTe2_Fig4}, we can be reassured that the irreversibility from the ZFC and FC curves from our $M$($T$) measurements is not originating from a spin-glass behavior, as we do not observe any frequency dependence in the AC susceptibility. Looking at only the AC susceptibility, a clear difference is seen depending on the orientation of the crystal to the applied AC field. For the \textbf{H} $\parallel$ \textbf{ab} orientation: (i) The unusual peak in the AC susceptibility around 130 K is more pronounced. (ii) The imaginary part (see inset) has a higher susceptibility above $\approx$ 124 K for higher frequencies that switches to a lower susceptibility below it and shows an additional shoulder at 130 K. The real part  always shows higher susceptibility for the lower frequencies. (iii) Below T$^*$ the system shows a negligible dissipation.
For the \textbf{H} $\perp$ \textbf{ab} orientation: (i) The unusual peak in the AC susceptibility around 130 K is less pronounced. (ii) The imaginary part  always has a higher AC susceptibility for higher frequencies and vice versa, and (iii) The imaginary part increases with decreasing temperature below T$_{\rm C}$ and for increasing magnetic field frequencies. 

The anomaly reflected by the unusual peak in the AC susceptibility above $T_{\rm C}$ has previously been observed for a different 2D magnetic system, i.e. \ce{VI3}, and was ascribed to surface layers that are suffering from lattice defects or iodine deficiency mimicking intralayer tensile strains.\cite{valenta2021pressure} Considering that our ZFC and FC measurements merge above the $T_{\rm C}$ of 117 K (see Fig. \ref{KCrTe2_Fig4} (c, e) and the mobility of the K cation, a similar scenario of tensile strain could likely be possible in this K-deficient system.

\subsection*{Muon spin spectroscopy}

\begin{figure*}
\begin{center}
\includegraphics[width=\linewidth]{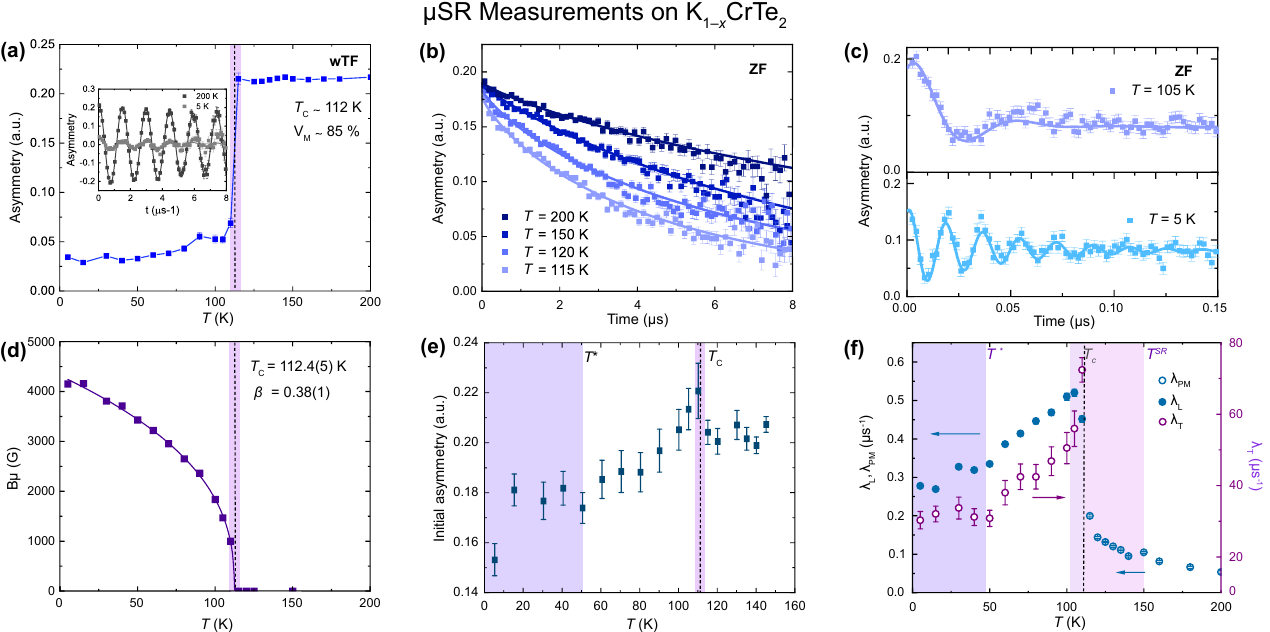}
\caption{\textbf{$\mu$SR measurements on \ce{K_{1-x}CrTe2} single crystals.} (a) Temperature dependence of the wTF
asymmetry and the resulting transition temperature. Inset: wTF spectra in the long time domain above and bellow the transition temperature of \ce{K_{1-x}CrTe2}. (b) ZF-\(\mu\)SR spectra above and (c) below the magnetic ordering temperature. Lines show the fitting to Eq. \ref{eq_muSR-stretch} and \ref{eqmuSR_ZF} respectively, see the text for details. (d) Temperature dependence of the internal field (\(B_\mu\)). The line shows the fitting to a power law with a critical exponent of \(\beta\)= 0.38(1). (e) Temperature dependence of the initial asymmetry. The shaded regions indicate the critical temperatures from magnetization measurements (\textit{T}$^{SR}$ = 150 K, \textit{T}$_{\rm C}$ = 117 K, \textit{T}$^*$ $\approx$ 50 K). (f) Temperature dependence of the muon spin relaxation rates \(\lambda\)\(_{pm}\) (blue circles), \(\lambda\)\(_{L}\) (filled purple circles, left axis) and \(\lambda\)\(_{T}\) (open purple circles, right axis). }
\label{Figure5}
\end{center}
\end{figure*}

To gain further microscopic insight into the magnetic order and spin dynamics, we performed $\mu$SR measurements on \ce{K_{1-x}CrTe2} single crystals. 
In order to determine the magnetically ordered volume fraction, we used in a first step weak transverse field (wTF)-$\mu$SR measurements in an applied magnetic field of 5 mT, oriented perpendicular to the initial muon spin polarization. In this configuration, the amplitude of the oscillatory signal directly reflects the paramagnetic volume fraction of the sample. As shown in Fig. \ref{Figure5}(a), the oscillation amplitude exhibits a sharp decrease below $\approx$ 112 K, indicating the onset of magnetic order. From this reduction, we estimate that approximately 85${\%}$ of the sample volume becomes magnetically ordered. The residual oscillatory signal persisting below the magnetic transition temperature suggests that a paramagnetic contribution remains down to the lowest temperatures. This residual component likely originates from a combination of background signal and intrinsic sample inhomogeneities, such as defects and/or paramagnetic impurities. Similar behavior has been reported previously in related compounds, including LiCrTe$_{2}$ and NaCrTe$_{2}$.\cite{nocerino2024cr} 

In a second step, in order to get additional details about the magnetic state, such as internal field strength and the depolarization rates, zero field (ZF)-$\mu$SR data were recorded in the normal and magnetically ordered states and are shown in Figs. \ref{Figure5}(b) and (c), respectively. In the temperature range from 200 K down to 115 K, the $\mu$SR signal exhibits a pronounced relaxation that cannot be adequately described by a single exponential decay. Instead, the spectra are well captured by a stretched-exponential relaxation function with a stretching exponent $\beta$ $\simeq$ 0.85(5) according to Eq. \ref{eq_muSR-stretch}, suggesting a distribution of local magnetic environments and/or fluctuation rates, consistent with spatially inhomogeneous or short-range magnetic correlations:  
\begin{equation} \label{eq_muSR-stretch}
A(t) = A_0 \cdot exp\big[-(\lambda_{PM}\cdot t)^\beta] 
\end{equation}

where $t$ is the time after muon implantation, $A(t)$ is the time-dependent asymmetry and $A_0$ is the initial asymmetry. The substantial paramagnetic relaxation rate ($\lambda_{PM}$) indicates precursor magnetic correlations developing in this regime (see Fig. \ref{Figure5}(f)). Upon cooling below 115 K, the ZF-$\mu$SR spectra develop well-resolved oscillations, as exemplified in Fig. \ref{Figure5}(c), which are a hallmark of static long-range magnetic order. The presence of a single cosine precession demonstrates that the muon ensemble experiences a well-defined internal magnetic field consistent with a collinear ferromagnetic state, and thus supports the collinear ferromagnetic state determined from the NPD refinement. The ZF-$\mu$SR data can be fitted with a Lorentzian damped internal field function:
\begin{equation} \label{eqmuSR_ZF}
A_\mathrm{ZF}(t) = A_0\big[f\mathrm{_{osc}\cdot cos}\left(\gamma_\mu  B_\mathrm{int}\cdot t + \phi \right)\cdot \mathrm{exp}\left(-\lambda_T \cdot  t\right) + \left(1-f_\mathrm{osc}\right)\cdot \mathrm{exp}\left(-\lambda_L \cdot t\right)\big] 
\end{equation}

where $f_\mathrm{osc}$ is the oscillating fraction that arises from the component of muon-spins inside the sample that are perpendicular to the internal magnetic field $B_\mathrm{int}$, $\phi$ is a phase offset and $\gamma_\mu/(2\pi)~=~135.5$~MHz/T is the gyromagnetic ratio of the muon.
The depolarization rates below the ordering temperature $\lambda_{T}$ and $\lambda_{L}$ characterize the damping of the oscillating and non-oscillating part of the $\mu$SR signal respectively, and are plotted in Fig. \ref{Figure5}(f).
The damping $\lambda_{T}$ arises predominantly due to a distribution of magnetic fields at the muon site, caused by static disorder. The longitudinal relaxation rate $\lambda_{L}$ reflects relaxation due to dynamic fluctuations.

The temperature evolution of the fit parameters are shown in Figs. \ref{Figure5}(d-f), and are typical for the onset of a second order phase transition below $T_\mathrm{C} = 112.4(5)$~K. The temperature dependence of the oscillatory fraction is shown in Fig. \ref{SM_T-dependence_osc-fra} of the SM. The magnetically ordered fraction sharply increases below $T_\mathrm{C}$.
The internal field $B_{int}$, calculated from the muon precession frequency \(\omega\) as $B_{int}$ = \(\omega\)/\(\gamma_\mu\), can be considered as an order parameter in close relation to the internal magnetization. As shown in Fig. \ref{Figure5}(d), the internal field gradually increases below $T_\mathrm{C}$ reaching a saturation of 410 mT at base temperature.

The temperature dependence of $B_{int}$ follows indeed a power law behavior of the form $M \propto (T_{\rm C} - T)^\beta$. We obtain from this fitting a critical exponent of $\beta$ = 0.38(1) below \textit{T}$_{\rm C}$ (\(\mu\)SR) = 112.4(5) K. The obtained critical temperature agrees with the \textit{T}$_{\rm C}$ derived from the wTF-\(\mu\)SR spectra, and is in qualitative agreement with DC magnetization, where \textit{T}$_{\rm C}$ = 117 K was obtained. The determined critical exponent suggests a 3D Heisenberg universality class for \ce{K_{1-x}CrTe2}, as already observed for the related \ce{LiCrTe2} and \ce{NaCrTe2}.\cite{Nocerino2022, nocerino2024cr} 

The temperature dependence of the static and dynamic relaxation rates is shown in Fig. \ref{Figure5}(f). The transverse depolarization rate $\lambda_{T}$ approximately 70 times larger than the longitudinal (dynamic) relaxation rate $\lambda_{L}$, implying that the former is dominated by static contributions. Upon cooling, $\lambda_{L}$ increases and reaches a pronounced maximum at 112 K, indicative of a slowing down of magnetic fluctuations as the system approaches the magnetic transition. With further cooling, $\lambda_{L}$ decreases and eventually saturates below 50 K, signaling the freezing of spin dynamics and the establishment of a quasi-static or static magnetically ordered state. A large static depolarization rate with a finite dynamic component at base temperature indicates a predominantly static magnetic state with residual slow spin dynamics.

Regarding the total initial asymmetry, we observe a continuous decrease of the initial asymmetry from approximately 0.20 to 0.18 upon cooling from 112 K to 50 K, below which it saturates (see Fig. \ref{Figure5} (e)). This gradual loss of asymmetry signifies that a fraction of the sample experiences strong and broadly distributed internal magnetic fields, leading to muon-spin depolarization on timescales too fast to be detected. Such behavior is characteristic of disordered or spatially inhomogeneous magnetism, with the missing asymmetry corresponding to approximately 15${\%}$ of the sample volume residing in a highly disordered magnetic state.

Overall, the ${\mu}$SR response reveals a predominantly homogeneous magnetically ordered state, which constitutes the major fraction of the sample, coexisting with a highly disordered magnetic component that occupies approximately 15${\%}$ of the volume, as inferred from the missing initial asymmetry. In addition, a finite dynamic relaxation rate persists down to the lowest measured temperatures, indicating the presence of residual slow spin dynamics in the system. At present, it is not possible to unambiguously determine whether these residual spin fluctuations originate from the dominant ordered fraction or from the minority disordered component or from non-magnetic sample regions and background. One possible scenario is that the major ordered phase is fully static, while the minor disordered fraction retains slow, fluctuating magnetic moments. Alternatively, weak dynamics could be intrinsic to the ordered state itself, for example, due to low-energy collective excitations or domain-wall motion. However, the similar temperature evolution of the dynamic relaxation rate $\lambda_{L}$ and the loss of initial asymmetry provide an important constraint. Both quantities exhibit closely correlated temperature dependences across the magnetic transition, strongly suggesting a common microscopic origin. This correlation makes it most likely that the observed residual slow spin dynamics is associated primarily with the disordered and non-magnetic background fractions, rather than with the homogeneous ordered state.

\section*{Discussion}

With the combination of different used experimental techniques, we find \ce{K_{1-x}CrTe2} (with \textit{x} $\approx$ 0.3) single crystals to exhibit a magnetic long-range order at a Curie temperature of ~$T_{\rm C}$ = 117 K. The NPD data confirms that the system has the easy axis in-plane, which goes hand in hand with the results of the DFT calculations and the observations from DC magnetometry. That the layers couple ferromagnetically and the spins lie in-plane is in contrast to the other intercalated members of the family \textit{A}$_x$\ce{CrTe2} with \textit{A} = Li, Na, which display A-type AFM with the easy axis perpendicular to the respective planes. While all the members of this family have FM in-plane interactions in-line with the GKA rules, it is interesting to note that crystals of \ce{CrTe2} and the here presented \ce{K_{1-x}CrTe2} have their easy axis in-plane.\cite{roseler2025efficient} DFT calculations reveal that not all in-plane interactions for \ce{K_{0.64}CrTe2} are ferromagnetic, in contrast to the calculations done on \ce{CrTe2}. \cite{roseler2025efficient} This could be a result of contributions of additional exchange paths via alkali orbitals or due to longer intralayer distances because of the K-intercalation, see Table \ref{table1}, emphasizing that this compound lies near a crossover regime in the \ce{\textit{A}_{\textit{x}}Cr\textit{X}2} series’ magnetic phase diagram. The interlayer coupling of the intercalated members is already small due to the 2D nature of the crystal structure, and is further reduced with the size of the intercalants. As such, we can expect a different mechanism behind the experimental observation of the orientation of the spin for \ce{CrTe2}, which exhibits a strong interlayer coupling, and \ce{K_{1-x}CrTe2}.\cite{roseler2025efficient}

In summary, we have successfully synthesized and characterized a  member of the alkali-metal intercalated \ce{CrTe2} series as large single crystals using a K/Te self-flux. \ce{K_{1-x}CrTe2} crystallizes in the \textit{Cm} monoclinic space group with a non equilateral triangular lattice. The deficiency of the intercalants results in the alkali metal being coordinated trigonal prismatically rather than octahedrally, which is defined as "type I" $AMX_2$. NPD data reveals the spins to lie in-plane along either \textbf{a} or \textbf{b}. This is in line with our magnetization measurements, where \ce{K_{1-x}CrTe2} undergoes a transition towards ferromagnetic long-range order below $T_{\rm C}$ = 117 K with a saturation of 2.7 $\mu_B$ at a field of 1 T when measured parallel to the plane. Fitting of the temperature dependence of the internal magnetic field derived from the ZF-$\mu$SR to the power law $M \propto (T_{\rm C}-T)^\beta$ results in the critical exponent $\beta$ $\approx$ 0.38(1), close to the predicted critical exponent of 0.337 of the 3D Heisenberg class. Our dynamic susceptibility measurements show no spin-glass behavior but an unusual peak above $T_{\rm C}$ that could originate from surface defects as observed in other 2D magnetic systems.

\bibliography{scibib}
\bibliographystyle{ScienceAdvances}
%\bibliographystyle{Science}

%%%%%%%%%%%%%%%% ACKNOWLEDGEMENTS %%%%%%%%%%%%%%%

\section*{Acknowledgments}
%Here you can thank helpful colleagues who did not meet the journal's authorship criteria, or provide other acknowledgements that don't fit the (compulsory) subheadings below. Formatting requirements for each of these sections differ between the \textit{Science}-family journals; consult the instructions to authors on the journal website for full details. For example: F.~A. was funded by the Generous Science Agency grant~2372.
The authors thank Enrico Giannini for helpful discussions about the magnetic measurements.
\paragraph*{Funding:}
%List the grants, fellowships etc. that funded the research; use initials to specify which author(s) were supported by each source. Include grant numbers when appropriate or required by the funding agency.
C.W., F.E. and F.O.v.R were supported by the Swiss National Science Foundation (SNSF) under Grants No. PCEFP2\_194183 and No. 200021-204065. S.A.L acknowledges support from the Danish National Committee for Research Infrastructure through the ESS-Lighthouse Q-MAT and the Novo Nordisk Foundation under Grant No. NNF23OC0087229. H.O.J. acknowledges support through JSPS KAKENHI Grants No.~24H01668 and No.~25K08460. Part of the computation in this work has been done using the facilities of the Supercomputer Center, the Institute for Solid State Physics, the University of Tokyo. 
%Z.G. acknowledges support from the SNSF through SNSF Starting Grant No. TMSGI2\_211750.
M.M. was supported by the Swedish Research Council (VR, Dnr:s. 2021-06157, 2022-03936, 2025-07622 and 2025-08127) and the Swedish Foundation for Strategic Research (SSF) through SwedNess.

\paragraph*{Author contributions:}
F.O.v.R. and C.W. designed the experiments. C.W. synthesized the crystals and conducted and analyzed the magnetization experiments. F.E. solved the crystal structures and conducted the temperature-dependent SXRD experiments. S.L. and J.A.K. performed the $\mu$SR experiment and analyzed the data with contributions of Z.G.. V.P. and C.W. performed the NPD experiments and analyzed the data. H.O.J. conducted the DFT and Monte Carlo calculations. All authors contributed to the analysis of the data. F.O.v.R. and C.W. wrote the manuscript with contributions from all the authors.
\paragraph*{Competing interests:}
%Disclose any potential conflicts of interest for all authors, such as patent applications, additional affiliations, consultancies, financial relationships etc. See the journal editorial policies web page for types of competing interest that must be declared. If there are no competing interests, state:
The authors declare that they have no competing interests.
\paragraph*{Data and materials availability:}
%Specify where the data, software, physical samples, simulation outputs or other materials underlying the paper are archived. They must be publicly accessible when the paper is published (without embargo) and enable readers to reproduce all the results in the paper. Contact the editor if you’re unsure what needs to be shared.

%Our preference is for digital material to be deposited in a suitable non-profit online data or software repository that issues the material with a DOI. Alternatively, an institutional repository, subject-based archive, commercial repository etc. is acceptable, as are (short) supplementary tables or a machine-readable supplementary data file. ‘Available on request’ or personal web pages are not allowed.
The data needed to evaluate the conclusions in the paper are present in the paper and/or the Supplementary Materials and have been deposited to Zenobo under the following accession DOI: 10.5281/zenodo.20291736.
Deposition Numbers 2543556, 2543557, 2543558 contain the supplementary crystallographic data for this paper. These data can be obtained free of charge via the joint Cambridge Crystallographic Data Centre (CCSD) and Fachinformationszentrum Karlsruhe Access Structure service.
%Cite the relevant DOI \cite{dataset}, URL \cite{example_url} or reference \cite{example2} in this statement.
%These \textit{do not} count towards the reference limit if they are only cited in the acknowledgements. Be specific and state a unique identifier -- such as an accession number, software version number or observation ID -- so readers can easily retrieve the exact material used.

%Declare any restrictions on sharing or re-use -- such as a Materials Transfer Agreement (MTA) or legal restrictions -- which must be approved by your editor. Unreasonable restrictions will preclude publication. Consult the journal's editorial policies web page for more details.

%%%%%%%%%%%%%%%% SUPPLEMENT LIST %%%%%%%%%%%%%%%

% List the contents of your Supplementary Materials, including the numbers of any
% supplementary figures, tables, external data files etc. and any references that are
% cited only in the supplement. In this example, refs. 7-8 are cited only in the supplement.
% Fill out your numbers accordingly and delete any lines that aren't applicable.
\subsection*{Supplementary materials}
Materials and Methods\\
Supplementary Text\\
Figs. S1 to S3\\
Tables S1 to S5\\
References \textit{(7-\arabic{enumiv})}\\ % automatically fills out the last reference number
% (filling out the other numbers automatically is possible but fiddly and liable to break)

%%%%%%%%%%%%%%%% END OF MAIN TEXT %%%%%%%%%%%%%%%

\newpage

%%%%%%%%%%%%%%%% START OF SUPPLEMENT %%%%%%%%%%%%%%%

% Figures, tables, equations and pages in the supplement are numbered S1, S2 etc.
\renewcommand{\thefigure}{S\arabic{figure}}
\renewcommand{\thetable}{S\arabic{table}}
\renewcommand{\theequation}{S\arabic{equation}}
\renewcommand{\thepage}{S\arabic{page}}
\setcounter{figure}{0}
\setcounter{table}{0}
\setcounter{equation}{0}
\setcounter{page}{1} % not 0 as \newpage already started a supplementary page
% References continue the numbering from the main text.

%%%%%%%%%%%%%%%% SUPPLEMENT TITLE PAGE %%%%%%%%%%%%%%%

\begin{center}
\section*{Supplementary Materials for\\ \scititle}

% Author list for the supplement
% Indicate the corresponding authors, but do NOT include institutions here
% It would be nice if the template auto-generated this, but doing so is complicated...
\author
Catherine Witteveen,$^{\ast}$ 
Felix Eder, 
Sara A. L\'{o}pez-Paz,
Vladimir Pomjakushin,
Jonas A. Krieger,
Zurab Guguchia,
Harald O. Jeschke,
Martin M\aa nsson,
Fabian O. von Rohr

\small$^\ast$Corresponding author. Email: c.witteveen@ifw-dresden.de\\
\end{center}

% Fill out the numbers for each type of supplementary material,
% and delete any lines that aren't applicable.
% These are just example numbers that don't match the rest of this template.
\subsubsection*{This PDF file includes:}
Materials and Methods\\
Supplementary Text\\
Figures S1 to S3\\
Tables S1 to S5\\

\newpage

%%%%%%%%%%%%%%%% MATERIALS AND METHODS %%%%%%%%%%%%%%%

\subsection*{Materials and Methods}
\subsubsection*{Synthesis}

Single crystals of K\textsubscript{1--\textit{x}}CrTe\textsubscript{2} (\textit{x} $\approx$ 0.3) were prepared using a mixed K/Te self-flux.
Potassium (block, Sigma Aldrich, 99\%), chromium (powder, Alfa Aesar, 99.99\%), and tellurium (pieces, Alfa Aesar, 99.999\%) were used as received and placed in the molar ratios 6:1:8 (total mass 1.9 g) in an alumina Canfield crucible set consisting of a bottom and top crucible separated by a frit-disc.\cite{Canfield2016} All handling of educts and reaction products was done in an argon-filled glovebox. 
This assembly was subsequently sealed in a quartz ampule under dynamic vacuum after having been purged three times with Argon.
The ampules were then heated in a muffle furnace (heating rate 30 °C/h) to 1000 °C, and slowly cooled to 550 °C over four days (96h). 
The ampules were then immediately centrifuged to separate the crystals from the flux, resulting in large single crystals of up to 6 mm x 6 x mm x 0.5 mm and a silvery metallic luster. It should be noted that the quartz ampule walls tend to crack on the surface after centrifugation.

\subsubsection*{Energy dispersive X-ray spectroscopy}
The composition of the as-grown crystals from three different batches was analyzed using energy dispersive X-ray spectroscopy (EDX) on a JEOL JSM-7600F scanning electron microscope (SEM) with an accelerating voltage of 15 kV equipped with an Oxford Instruments X-Max 80 detector. The crystals were prepared onto the sample holders in the glovebox by covering them with kapton foil and securing the foil with kapton tape. The tape and foil were removed right before the transfer into the chamber. Details on the resulting composition of three different synthesized batches are collated in Tab. \ref{SM_EDX}

\begin{table}
	\begin{center}
	\caption{\textbf{Composition of various crystals determined by SEM-EDS.}}\label{SM_EDX}\label{SM_tab1}
		\begin{tabular}{ccccc}	
 	& spots & Cr:Te ratio & K-content & $\sigma$ (K)	\\
\hline
\multicolumn{5}{l} {Batch 1} \\
site 1 & 3 & 1.935 & 0.664 & 0.11 \\
\hline
\multicolumn{5}{l} {Batch 2} \\
site 1 & 4 & 1.931 & 0.749 & 0.06 \\
\hline
\multicolumn{5}{l} {Batch 3} \\
site 1 & 9 & 2.039 & 0.785 & 0.05 \\
\hline	\end{tabular}
	\end{center}
\end{table}
\renewcommand\baselinestretch{1.5}\selectfont

\subsubsection*{Single crystal X-ray diffraction}

The crystal structure of of K\textsubscript{1--\textit{x}}CrTe\textsubscript{2} was solved by SXRD on a Rigaku Supernova diffractometer using Mo-K\textsubscript{$\alpha$} radiation at 100 K. 
Unit-cell indexation, integration, and spherical absorption correction were performed with CrysalisPro \cite{CrysAlisPRO2022}. Small crystal pieces were cut using a scalpel in the glovebox and mixed with paraffin oil before being picked under an optical microscope and subsequently measured under the cold nitrogen gas flow. Details of the crystal structure determination can be found in Tab. \ref{SM_tab1}
Temperature-dependent SXRD measurements were performed on a different crystal but from the same batch in a range of 90–-280 K. For the redeterminations of the crystal structures of self-flux grown \ce{LiCrTe2} and \ce{NaCrTe2}, a similar procedure with measurement temperatures close to 100 K was followed.

\renewcommand\baselinestretch{1}\selectfont
\begin{table}
	\begin{center}
	\caption{\textbf{Details on crystal structure determination and refinement.}}\label{SM_tab2}

\begin{tabular}{llll}
 & & & \\
\hline
Refined Chemical formula 	&	 K\textsubscript{0.64(4)}CrTe\textsubscript{2} & LiCrTe\textsubscript{2} & NaCrTe\textsubscript{2}	\\
Molar mass (g*mol$^{-1}$) & 332.31 & 314.14 & 330.19\\
$\rho$\textit{\textsubscript{calc.}} (g*cm$^{-3}$) & 4.946 & 5.904 & 5.521\\
Temperature (K) & 100 & 90 & 100 \\
Space group & \textit{Cm} & $P\bar{3}m$ & $P\bar{3}m$\\
Space group no. & 8 & 164 & 164 \\
\textit{a} ($\mathring{\mathrm{A}}$) & 6.7539(4) & 3.96730(10) & 3.99790(10) \\
\textit{b} ($\mathring{\mathrm{A}}$) & 3.9466(3) & & \\
\textit{c} ($\mathring{\mathrm{A}}$) & 8.6549(7) & 6.58760(10) & 7.3735(2) \\
\textit{$\beta$} (°) & 104.705(8) & & \\
\textit{V} ($\mathring{\mathrm{A}}$$^3$) &  223.14(3) & 89.794(5) & 102.063(6) \\
\textit{Z} & 2 & 1 & 1 \\
Crystal color & black & black & grey \\
Crystal shape & block & plate & bar \\
Crystal size (mm$^3$) & 0.27$\times$0.18$\times$0.11 & 0.25$\times$0.11$\times$0.02 & 0.18$\times$0.04$\times$0.02 \\
Absorption correction & \multicolumn{3}{c}{spherical} \\
$\mu$ (mm\textsuperscript{--1}) & 23.96 & 18.82 & 16.67\\
Diffractometer & \multicolumn{3}{c}{Oxford Diffraction SuperNova} \\
Radiation; $\lambda$ ($\mathring{\mathrm{A}}$) & \multicolumn{3}{c}{MoK\textsubscript{$\alpha$}; 0.71073} \\
Reflections used & 5709 & 6662 & 8085\\
$\theta$\textsubscript{min} -- $\theta$\textsubscript{max} (°) & 4.870 --  29.129 & 3.092 -- 34.118 & 2.762 -- 34.853 \\
\textit{h} range & --9 to 9 & --6 to 6 & --6 to 6\\
\textit{k} range & --5 to 5 & --6 to 6 & --6 to 6\\
\textit{l} range & --11 to 11 & --10 to 10 & --11 to 11\\
\textit{R\textsubscript{int}} & 0.066 & 0.083 & 0.067 \\
\textit{R\textsubscript{$\sigma$}} & 0.025 & 0.015 & 0.014 \\
Independent refl. & 688 & 175 & 207 \\
Observed refl. (\textit{I} $<$ 3$\sigma$(\textit{I})) & 666 & 171 & 202 \\
\textit{R1} (\textit{I} $<$ 3$\sigma$(\textit{I})); (\%) & 3.62 & 2.71 & 1.73 \\ 
\textit{wR2\textsubscript{all}} (\%) & 11.51 & 6.99 & 4.52 \\
Goodness of fit & 1.152 & 1.137 & 1.174 \\
Maximal difference peaks (\textit{e}\textsuperscript{--}$\mathring{\mathrm{A}}$\textsuperscript{--3}) & --1.44; 2.56 & --1.69; 2.90 & --1.15; 0.98 \\
CCDC deposition code & 2543556 & 2543557 & 2543558 \\
\hline
	\end{tabular}
	\end{center}
\end{table}

\renewcommand\baselinestretch{1.5}\selectfont

\subsubsection*{Magnetization measurements}
The temperature-dependent and field-dependent direct current (DC) magnetization was measured in a Physical Property Measurement System (Quantum Design PPMS DynaCool) equipped with a 9 T magnet with the vibrating sample magnetometer (VSM) option. The temperature-dependent alternating current (AC) magnetization was measured on the same Quantum Design PPMS DynaCool with the AC measurement system (ACMS) option, for both orientations and using a 10 Oe AC drive across a range of four frequencies.
Zero field cooled (ZFC) magnetic moment data was collected upon cooling without field and measuring while heating, while field cooled (FC) data was collected upon cooling in the applied field and measuring while heating at the given field strengths. Crystals from two different batches were measured, showing the same response. Sample preparation and mounting were done in the glovebox. Crystals were coated in N-apiezon grease and inserted in a Kapton envelope made from Kapton foil and Kapton tape which were subsequently mounted on the sample holder to avoid decomposition. The data was not corrected for the demagnetizing factor. 

\subsubsection*{Neutron powder diffraction experiments}
This work is based on experiments performed at the Swiss spallation neutron source SINQ, Paul Scherrer Institute (PSI), Villigen, Switzerland.
Neutron diffraction patterns were obtained on the High-Resolution Powder Diffractometer at the SINQ from PSI.\cite{fischer2000_HRPT} Data were collected at \textit{T} = 1.8 K and 150 K with a neutron wavelength of 1.886 $\mathring{\mathrm{A}}$ for magnetic structure refinements and structural refinements (150 K) respectively and analyzed by the Rietveld method using the Fullprof Suite package.\cite{Fullprof} The magnetic symmetry analysis was done using ISODISTORT from the ISOTROPY software and BasIreps from the Fullprof suite.\cite{ISODISTORT} The peak shape was modeled using a Thompson–Cox–Hastings pseudo-Voigt function with axial divergence asymmetry (as implemented in Fullprof with Npr = 7). Around 380 mg Single crystals were finely powderized in an argon filled glovebox using an agate mortar. The powder was then transferred to a vanadium sample container of 6 mm diameter and 100 mm length and sealed with an indium ring under helium atmosphere. The secondary phase observed in the pattern belongs to vanadium from the vanadium container. The SXRD data for K\textsubscript{0.64}CrTe\textsubscript{2}listed in table \ref{SM_tab2} was used as a starting model. The atomic coordinates and potassium content were kept at fixed values.
Table \ref{SM_Table3} shows the structural parameters obtained from Rietveld refinements of the NPD data ($\lambda$ = 1.886 $\mathring{\mathrm{A}}$) at \textit{T} = 150 K and \textit{T} = 1.8 K for K$_{0.64}$CrTe$_2$, together with the agreement factors for both temperatures and solutions.

\renewcommand\baselinestretch{1}\selectfont
\begin{table}
\centering
\caption{\textbf{Refined parameters obtained from NPD with the spacegroup \textit{Cm} (no. 8).}}
  \label{SM_Table3}
  \begin{tabular*}{\textwidth}{@{\extracolsep{\fill}}  c c c c c}
    \hline
                &              & \textbf{150 K}       & \textbf{1.8 K a }     & \textbf{1.8 K b }    \\
    \hline
    
          & \textit{a} ($\mathring{\mathrm{A}}$)     & 6.7425(8)  & 6.7525(5)  & 6.7611(4)\\
          & \textit{b} ($\mathring{\mathrm{A}}$)     & 3.9348(5)  & 3.9436(3)  & 3.9377(2)\\
         & \textit{c} ($\mathring{\mathrm{A}}$)     & 8.6757(10)  & 8.6367(6)  & 8.6432(5)\\
         & \textit{V} ($\mathring{\mathrm{A}}$ \textsuperscript{3})    &  222.750(48) & 222.558(30)  & 222.67(2)  \\
          & \textit{$\beta$} (°)    &  104.594(11) & 104.611(8)  & 104.611(6)  \\
                      \hline
           Atom          & Site; Wyckoff pos         &           &       \\
    \hline
           K            & (0.84390, 0, $\frac{1}{2}$); 2\textit{a}  &           &       \\
                        & occ. & 0.64            &   0.64 &   0.64     \\
                         & \textit{B}\textsubscript{\textit{iso}} ($\mathring{\mathrm{A}}$ \textsuperscript{3}) & 3.76(58) & 2.66(37) & 2.51(38)        \\
    \hline
           Cr            & (0, 0, 0); 2\textit{a}  &           &       \\
                        & occ. &   1        &   1        & 1     \\
                         & \textit{B}\textsubscript{\textit{iso}} ($\mathring{\mathrm{A}}$ \textsuperscript{3}) & 1.13(16)& 1.09(12)  & 1.19(13)       \\   
    \hline
           Te1           & (0.39517, 0,    0.18341); 2\textit{a}  &           &       \\
                         
                & occ.  &     1      &   1  &   1    \\
                         & \textit{B}\textsubscript{\textit{iso}} ($\mathring{\mathrm{A}}$ \textsuperscript{3}) & 1.25(54) & 0.97(43)  & 1.52(38)     \\
                          \hline
           Te2           & (0.60676, 0,    0.81808); 2\textit{a}  &           &       \\
                         
                & occ.  &     1      &    1 &    1     \\
            & \textit{B}\textsubscript{\textit{iso}} ($\mathring{\mathrm{A}}$ \textsuperscript{3}) & 1.45(54) & 0.41(38)  & 1.82143     \\
        \hline
                         & $\mu_{Cr}$           & -       & 2.74(2) & 2.72(3)\\
                         & $R_{B(cryst)}$ (\%)  & 7.76      & 11.4  & 10.1       \\
                         & $R_{B(mag)}$ (\%)    & -        & 7.35  & 6.46       \\
                         & $\chi\textsuperscript{2}$   &  2.09   & 4.04     & 4.53   \\
                         
    \hline
  \end{tabular*}
\end{table}
\renewcommand\baselinestretch{1.5}\selectfont

\newpage

\subsubsection*{Muon spin spectroscopy}
This work is based on experiments performed at the Swiss Muon Source S$\mu$S, Paul Scherrer Institute, Villigen, Switzerland.
Zero field (ZF) and weak transverse field (wTF) $\mu$SR experiments were carried out at the $\pi$M3 beam line (low background GPS instrument), using an intense beam ($p_\mu$ = 29 MeV/c) of positive muons.\cite{amato2017GPS} An envelope of aluminum foil was made in which several crystals were stacked, all inside a helium glovebox. There, the envelope was wrapped in kapton tape for air tight insulation and secured on the copper sample holder with mylar tape. Subsequently, it was mounted in a helium gas-flow cryostat. The temperature was varied between 5 and 200 K. All measurements were performed with the initial muon spin direction rotated 6° away from the beam direction. For wTF-$\mu$SR experiments the sample was aligned such that beam direction corresponded to the c-axis of the crystals. On the other hand, for ZF-$\mu$SR measurements, the c-axis was rotated 45° away from the beam, such that the muon spin precession was sensitive to both in-plane and out-of-plane local field components.

%%%%%%%%%%%%%%%% SUPPLEMENTARY TEXT %%%%%%%%%%%%%%%

\newpage

\subsubsection*{Supplementary note 1: DFT results}

In Tables~\ref{tab:couplings} and \ref{tab:anisotropies}, we provide the detailed results of the DFT energy mapping calculations. In Table~\ref{tab:couplings}, exchange interactions are given for $S=3/2$ and without double counting of bonds. The given errors are statistical errors of the fit. From energy differences between $x$, $y$ and $z$ quantization axes in fully relativistic DFT+U calculations, we extract the parameters of the anisotropy term
\begin{equation}
    H_{\rm aniso}=K_x S_x^2+K_z S_z^2\,,
\end{equation}
where the length of $S_x$ and $S_z$ is taken to be $3/2$. The results are listed in Table~\ref{tab:anisotropies}.

\begin{table}
\centering
\caption{\textbf{Exchange couplings of \ce{K_{0.64}CrTe2} calculated by DFT energy mapping.} The Hund's rule coupling was fixed at $J_{\rm H}=0.72$\,eV~\cite{Mizokawa1996}.}  
\begin{tabular}{c|c|c|c|c|c|c|c|c|c}
$U$\,(eV)&$J_1$\,(K)&$J_2$\,(K)&$J_3$\,(K)&$J_4$\,(K)&$J_5$\,(K)&$J_6$\,(K)&$J_7$\,(K)&$J_9$\,(K)&$\theta_{\rm CW}$\,(K)\\\hline
1.5 & --33.7(9) & --43.7(1.0) & 4.9(1.0) & 4.8(5) & 13.0(5) & 11.9(1.0) & --4.6(5) & --3.1(4) & 173 \\%# K0.64CrTe2, U=1.5, 4x4x4
1.75 & --36.3(8) & --45.9(9) & 4.2(9) & 4.1(5) & 12.2(5) & 11.4(9) & --4.1(5) & --2.9(4) & 201 \\%# K0.64CrTe2, U=1.75, 4x4x4
2 & --38.8(7) & --48.0(8) & 3.5(8) & 3.3(4) & 11.6(4) & 11.0(8) & --3.6(4) & --2.8(3) & 226 \\%# K0.64CrTe2, U=2, 4x4x4
2.25 & --41.2(7) & --50.1(7) & 2.8(7) & 2.6(4) & 11.0(4) & 10.6(7) & --3.2(4) & --2.7(3) & 251 \\%# K0.64CrTe2, U=2.25, 4x4x4
2.5 & --43.6(6) & --52.1(7) & 2.1(7) & 1.9(4) & 10.6(4) & 10.3(7) & --2.8(4) & --2.5(3) & 274 \\%# K0.64CrTe2, U=2.5, 4x4x4
2.75 & --45.9(5) & --54.0(6) & 1.6(6) & 1.3(3) & 10.3(3) & 10.2(6) & --2.4(3) & --2.3(2) & 295 \\%# K0.64CrTe2, U=2.75, 4x4x4
3 & --48.1(5) & --55.9(5) & 1.2(5) & 0.6(3) & 10.2(3) & 10.2(5) & --1.9(3) & --2.2(2) & 314 \\%# K0.64CrTe2, U=3, 4x4x4
\hline
\end{tabular}
      \label{tab:couplings}
\end{table}

\begin{table}
\centering
\caption{\textbf{Single ion anisotropies of \ce{K_{0.64}CrTe2} calculated calculated using fully relativistic DFT+U calculations.} The Hund's rule coupling was fixed at $J_{\rm H}=0.72$\,eV~\cite{Mizokawa1996}.} 
\begin{tabular}{c|c|c}
$U$\,(eV)&$K_x$\,(K)&$K_z$\,(K)\\\hline
1.5 & 0.27 & 4.43 \\
1.75 & 0.27 & 4.34 \\
2 & 0.27 & 4.29 \\
2.25 & 0.27 & 4.31 \\
2.5 & 0.27 & 4.38 \\
2.75 & 0.27 & 4.48 \\
3 & 0.27 & 4.61 \\
\hline
\end{tabular}
       \label{tab:anisotropies}
\end{table}

\subsubsection*{Supplementary note 2: Magnetization measurements}

\begin{figure}
\begin{center}
\includegraphics[width=\linewidth]{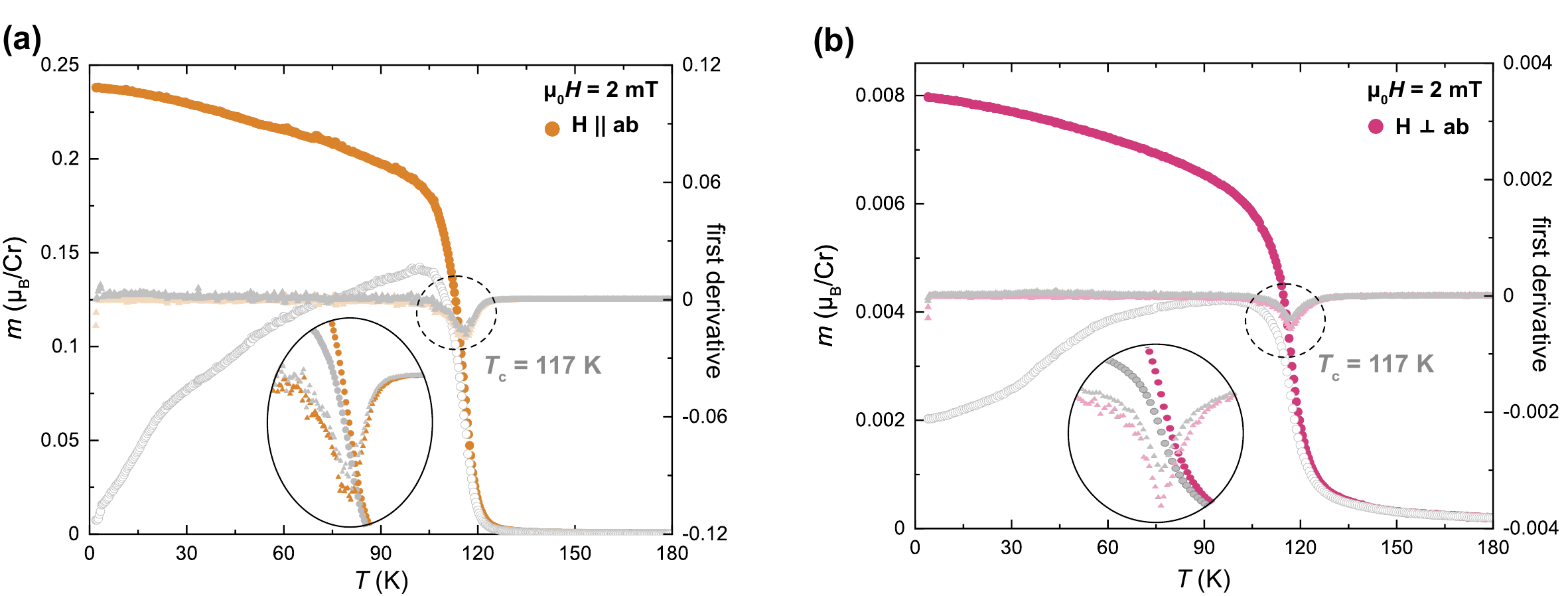}
\caption{\textbf{ZFC and FC DC magnetization and its derivative of the K$_{1-\textit{x}}$CrTe$_2$ single crystal of both orientations at an applied field of 2 mT.} (a) \textbf{H} $\parallel$ \textbf{ab} (orange) and (b) \textbf{H} $\perp$ \textbf{ab} (pink).} 
\label{Si_mag_derivative}
\end{center}
\end{figure}

The transition temperature was determined with the first derivative on the FC and ZFC data taken with an external field of 2 mT and resulted in a \textit{T}$_c$ of 116.85 K for both orientations, as can be seen in Fig. \ref{Si_mag_derivative}. The ZFC and FC data only merge shortly before 150 K. The insets show a zoom of the indicated regions marked as dashed circles.

\newpage
\subsubsection*{Supplementary note 3: NPD experiments}

\begin{figure}
\begin{center}
\includegraphics[width=0.7\linewidth]{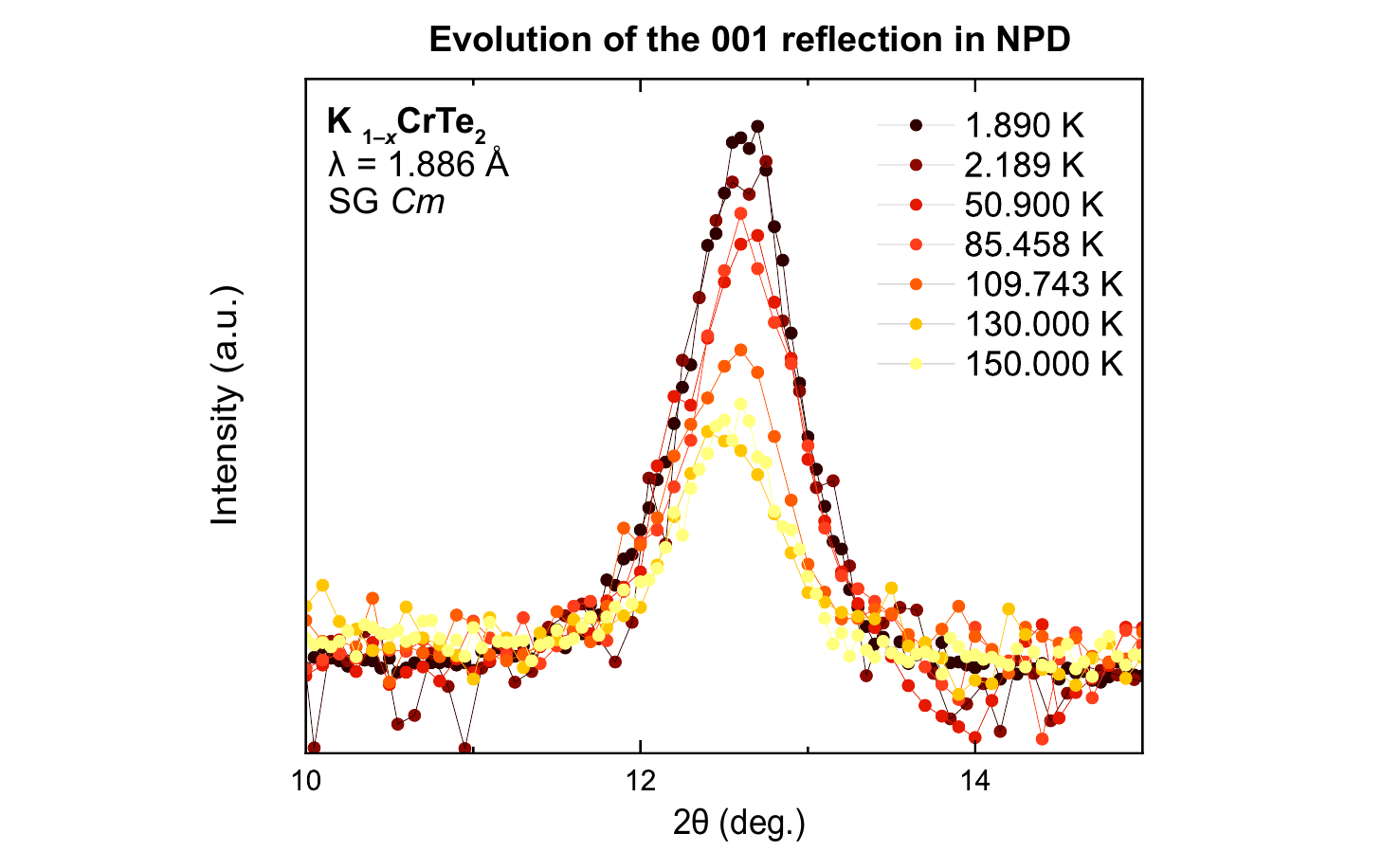}
\caption{\textbf{Evolution of the (001) reflection with temperature.}} 
\label{SM_NPD_T}
\end{center}
\end{figure}

\begin{figure}
\begin{center}
\includegraphics[width=0.5\linewidth]{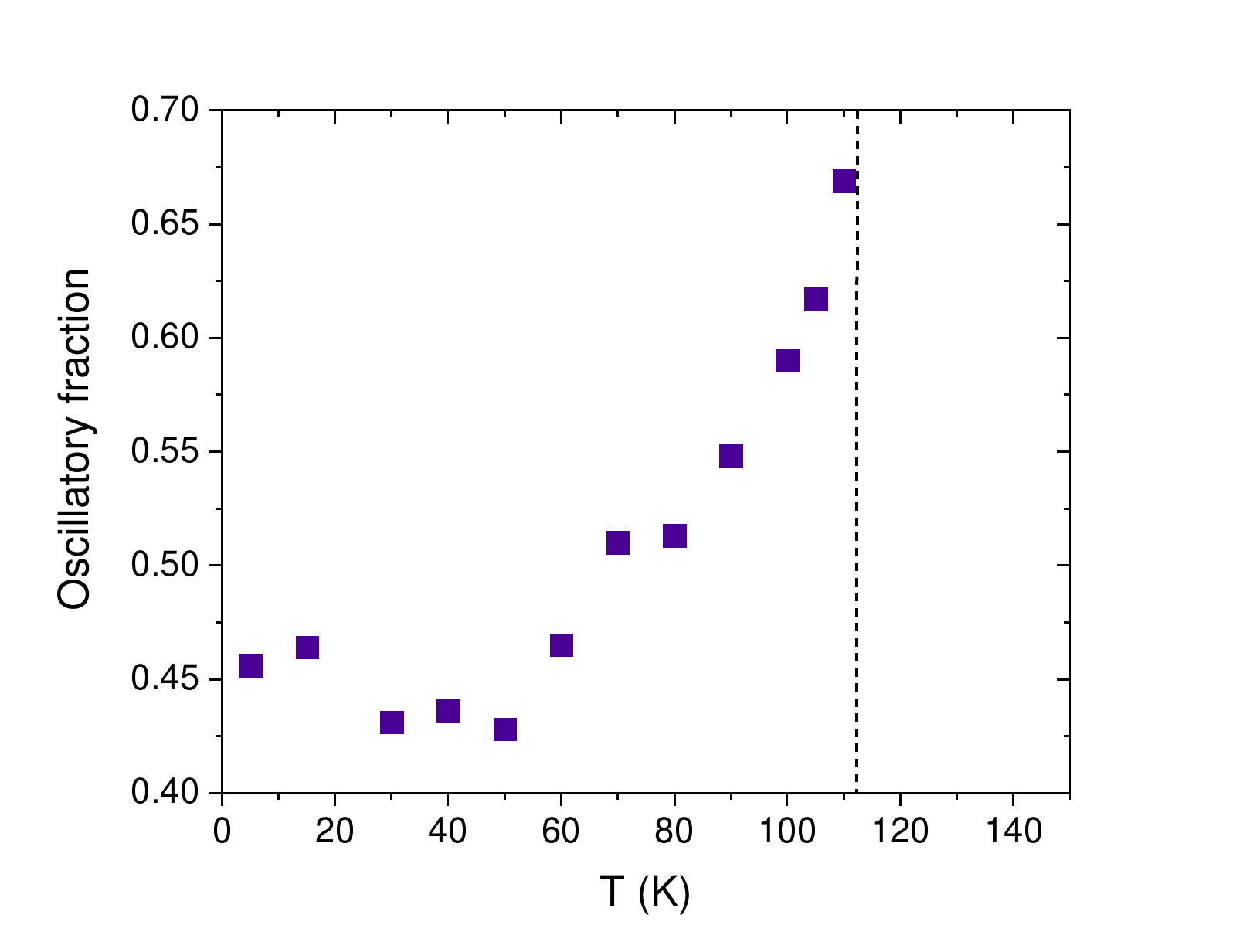}
\caption{\textbf{Temperature dependence of the oscillatory fraction} }
\label{SM_T-dependence_osc-fra}
\end{center}
\end{figure}

Figure \ref{SM_NPD_T} shows the evolution of the intensity of the (001) reflection with temperature. At 130 K and 150 K, the intensity remains the same, indicating no starting magnetic long-range order.

\newpage

\subsubsection*{Supplementary note 4: $\mu$SR experiments}

%%%%%%%%%%% CAPTIONS FOR OTHER SUPPLEMENTARY FILES %%%%%%%%%%

\clearpage % Clear all remaining figures and tables then start a new page

%%%%%%%%%%%%%%%% SUPPLEMENTARY REFERENCES %%%%%%%%%%%%%%%

% Do NOT include a reference list in the supplement.
% All references must be in a single list at the end of the main text.
% The copyeditors will ensure that the correct reference list appears with each version of the paper
% (print, HTML, PDF, mobile app, metadata for bibliographic databases etc.)

% After the paper has completed peer review and been revised ready for acceptance,
% you should comment out the lines above and copy-paste the contents of your .bbl
% file here instead. This will help ensure that our conversion software works correctly.
% Remember to re-run BibTeX first - check the timestamp!
%
% Example of the first three entries copy-pasted from science_template.bbl:
%
%\begin{thebibliography}{1}
%
%\bibitem{example}
%A.~N. {Author}, An example reference. \emph{Journal of Improbable Research}
%  \textbf{1}, 67 (2020).
%
%\bibitem{example2}
%F.~M. {Surname}, S.~{Author}, A second example. \emph{Interesting Research
%  Letters} \textbf{32}, 897 (2019).
%
%\bibitem{example_preprint}
%P.~{One}, P.~{Two}, P.~{Three}, {An unpublished preprint}. \emph{preprint}
%  (2021), arXiv:2101.12345.
%
%\end{thebibliography}

\end{document}